\documentclass[12pt]{article}
\usepackage{appendix}
\usepackage{anysize}
\usepackage{graphics}
\usepackage{color}
\usepackage{amssymb}
\usepackage{graphicx}
\usepackage{epstopdf}
\usepackage[hang,small,bf]{caption}
\usepackage{hyperref}
\usepackage{amsmath}
\usepackage{latexsym}
\usepackage{setspace}
\usepackage{sectsty}
\usepackage{breqn}

\usepackage{graphicx}
\usepackage{dcolumn}
\usepackage{bm}
\usepackage{hyperref}
\usepackage{placeins}

\def \beq{\begin{equation}}
\def \eeq{\end{equation}}
\def \bea{\begin{align}}
\def \eea{\end{align}}

\def\lsim{\mathrel{\rlap{\lower4pt\hbox{\hskip1pt$\sim$}}
    \raise1pt\hbox{$<$}}}                % less than or approx. symbol
\def\gsim{\mathrel{\rlap{\lower4pt\hbox{\hskip1pt$\sim$}}
    \raise1pt\hbox{$>$}}}                % greater than or approx. symbol

\title{
\vspace*{-1.3cm}
\begin{flushright}
\normalsize{
ANL-HEP-PR-11-46\\
EFI-11/22\\
FERMILAB-PUB-11-337-T\\
SCIPP 11/07}
\end{flushright}
\vspace*{0.5cm}
\Large
\textbf{The 7 TeV LHC Reach for MSSM Higgs Bosons}
\author{Marcela~Carena$^{1,2}$, Patrick~Draper$^{2,3}$, \\
  Tao~Liu$^{4}$ and Carlos~E.M.~Wagner$^{2,5,6}$ \\
~\\
\normalsize\emph{$^1$~Fermi National Accelerator Laboratory, P.~O.~Box 500,
  Batavia, IL 60510, USA}\\ 
\normalsize\emph{$^2$~Enrico Fermi Inst., Univ. of Chicago, 5640 S. Ellis Ave.,
  Chicago, IL 60637, USA}\\ 
  \normalsize\emph{$^2$~Santa Cruz Inst. for Particle Physics, Univ. of California, Santa Cruz, CA 95064, USA}\\
\normalsize\emph{$^4$~Department of Physics, Univ. of California, Santa Barbara, 
  CA 93106, USA}\\
\normalsize\emph{$^5$~HEP Division, Argonne National Laboratory, 9700 Cass
  Ave., Argonne, IL 60439, USA}\\ 
\normalsize\emph{$^6$~KICP and Dept. of Physics, Univ. of Chicago, 5640 S. Ellis
  Ave.,Chicago IL 60637, USA}}}
\begin{document}
\nocite{*}
\setcounter{page}{0}
\maketitle
\thispagestyle{empty}
%\vspace{-0.5cm}
%\date{\today}
\begin{abstract}
The search for the Higgs boson is entering a decisive phase. The Large Hadron Collider experiments have collected more than 1~fb$^{-1}$ of data and are now capable of efficiently probing
the high Higgs mass region, $m_H > 140$~GeV.  The low mass region is more challenging at the LHC, but if the Higgs 
has Standard Model (SM)-like
properties, the LHC should find evidence for it by the end of next year.
% At the same time, the Tevatron experiments have collected more than 10~fb$^{-1}$ and with improved analyses, should be able to probe the existence of a SM-like Higgs with mass below 180~GeV by the end of next year.  
In low energy supersymmetric extensions of the SM, the situation is similar for large values of the $CP$-odd Higgs mass $m_A$, but more interesting for lower values of $m_A$.  The ($\sqrt{s}$ =7 TeV) LHC searches for a low-mass Standard Model Higgs boson predominantly in the $h\rightarrow \gamma\gamma,WW$ decay modes, which may be suppressed by an increase in the $h\rightarrow b\bar{b}, \tau^+ \tau^-$ partial widths (and thus the total $h$ width) for $m_A\lesssim 500$ GeV. Although $h\rightarrow b\bar{b}, \tau^+ \tau^-$ are sought at the LHC, these channels are not powerful enough to fully counter this suppression in the first year of running. We consider two alternative possibilities for probing the low $m_A$ region: nonstandard Higgs boson  searches at the LHC, and a statistical combination with the Tevatron, where $Vh\rightarrow b\bar{b}$ is the primary search channel for light $h$. We also study an MSSM scenario in which the $h\rightarrow\gamma\gamma$ rate is enhanced at low $m_A$ to the point where discovery is possible in the near future.
%, while the Tevatron searches for a light Higgs boson primarily in . We demonstrate that the Tevatron and LHC search modes are complementary when interpreted on the Higgs sector parameter space of the Minimal Supersymmetric Standard Model: for large values of the $CP$-odd Higgs mass, the Tevatron is weakest because the Higgs is heavier, while for lower values of the $CP$-odd mass, the main LHC searches are suppressed by a large enhancement of the $h\rightarrow b\bar{b}$ partial width. We estimate the expected, statistically combined reach of the two colliders in the next few years, and show that it may provide evidence for an SM-like Higgs boson in regions where neither collider is powerful enough to do so alone. Finally, we show that nonstandard search channels are also complementary with SM-like channels at the LHC, and that a combination may probe much of the MSSM Higgs sector at the LHC alone.

\end{abstract}

\newpage
\setcounter{page}{1}

%\pacs{12.60.Fr, 13.20.Gd, 13.66.Hk, 14.80.Cp}

%\maketitle
%\renewcommand{\thepage}{\arabic{page}}

\section{Introduction}

The Standard Model (SM) provides a very good description of experimental observables measured at high-energy colliders.  The SM is a renormalizable theory and admits a perturbative description at scales of the order of the weak scale. The Higgs boson~\cite{Higgs:1964pj,Higgs:1966ev} is the only element of the SM that has not been discovered, and plays an important role in ensuring the perturbative consistency of the theory. Within the SM, precision electroweak observables suggest a light Higgs boson, with mass below about 180~GeV. Searches for a Higgs particle are therefore some of the most important activities in high energy physics. Currently, collider searches are performed at the LHC and the Tevatron experiments, and Tevatron data has already 
excluded the presence of a SM-like Higgs boson with a mass in the range 158--173~GeV at the 95\% confidence level~\cite{Aaltonen:2011gs}.  
%The mass region between 125 and 140~GeV becomes more challenging at the Tevatron, and will require analysis improvements, which are expected to be realized in the near future. 
At CERN, the LHC is accumulating record high luminosities, and it is expected to probe the whole SM Higgs mass region below~500 GeV by the end of 2011.  The most challenging mass region for Higgs searches at the LHC is the 
closest to the current LEP bound of about 115~GeV. 
%, where the Tevatron has somewhat enhanced capabilities.
In this low mass region, the main search channel at the LHC comes from the Higgs production via gluon fusion and its rare decay into two photons~\cite{Aad:2011qi,Aad:2009wy,Ball:2007zza,lhc2fc}. Other relevant search channels, which require higher statistics, are weak boson fusion with $h \to \tau^+\tau^-$ and Higgs associated production with weak vector bosons, with $h \to b \bar{b}$.
%, while at the Tevatron the Higgs is sought in associated production with weak gauge bosons and decays into two bottom quarks.
It is expected that by the end of the year the associated production $Vh\rightarrow b\bar{b}$ channels at the Tevatron will be able 
to test the SM-Higgs mass region close to the present LEP bound~\cite{cdfnote}.  

In this note we concentrate on searches for neutral Higgs bosons in the $CP$-conserving Minimal Supersymmetric Standard Model (MSSM)~\cite{reviews}.  In most of the MSSM parameter space there is a light Higgs boson
with SM-like couplings to gauge bosons and a mass below 130~GeV~\cite{mhiggsRG1a,mhiggsRG1,HHH,Degrassi:2002fi}.   Additional $CP$-even, $CP$-odd, and charged Higgs bosons exist in this model and possess enhanced
couplings to the third generation fermions~\cite{hhg}.  Searches for these non-standard Higgs bosons are being performed at the Tevatron and the LHC, with the LHC rapidly surpassing
the Tevatron capabilities in the main modes where neutral $CP$-even and $CP$-odd Higgs bosons decay into $\tau$-lepton pairs~\cite{Chatrchyan:2011nx,atlastautau}. 

In previous articles we have studied the reach of both the Tevatron and a 14~TeV LHC in their searches for standard and non-standard Higgs bosons of the MSSM~\cite{Draper:2009fh,Draper:2009au,Carena:2010ev}. 
Since the LHC is now operating at a center of mass energy of 7~TeV, it is important to perform a realistic estimate of its reach in the first years of running. In the course of this analysis we stress the fact that in supersymmetry, the presence of more than one Higgs doublet means that mixing between the neutral components can produce a state with SM-like gauge couplings but very different branching fractions from the SM Higgs boson.  In general, the $h\rightarrow b\bar{b}$ width of the SM-like Higgs boson is increased to a degree controlled by the $CP$-odd mass $m_A$, and this effect suppresses the rates into other states such as $h\rightarrow\gamma\gamma,WW$ (the diphoton suppression was also discussed in~\cite{Carena:1999bh}, and more recently in~\cite{Cao:2011pg}). For more specialized values of the soft supersymmetry-breaking parameters, the $h\rightarrow b\bar{b}$ width can be suppressed and the rates into other states are enhanced. We study both of these possibilities in detail and demonstrate that when the main LHC searches for $h$ are weakened, either LHC nonstandard Higgs searches can be used to probe the parameter space, or a statistical combination with the Tevatron data may be used to provide evidence for the presence of $h$. On the other hand, when $h\rightarrow\gamma\gamma$ is enhanced in the MSSM, we show that the LHC will quickly reach discovery potential for the SM-like Higgs boson, while non-standard Higgs searches will provide a complementary search channel. 

This article is organized as follows. In section 2, we discuss the statistical methods used in our analysis. In section 3 we show the LHC results in different benchmark
scenarios. In section 4 we demonstrate that non-standard Higgs boson searches, as well as searches for SM-like Higgs bosons at the Tevatron, offer power complementary 
to that of the SM-like Higgs channels at the LHC. In section 5 we conclude. 

% of these colliders at the end of the
%Tevatron run and in the first years of the LHC. In this article we present such estimates.

\section{Methods}

Searches for SM-like Higgs bosons at hadron colliders are performed in a diverse set of channels, and the reach of SM-like Higgs bosons searches has been thoroughly studied by the experimental collaborations. We base our analysis on the results of these experimental analyses, properly interpreting them in the MSSM context and combining the significance of different channels.

Since the reaches for SM-like Higgs bosons at CMS and ATLAS are quite similar, for simplicity, we estimate the combined LHC reach by doubling  the luminosity at the ATLAS experiment and assuming 5 fb$^{-1}$/channel/experiment  (i.e., all channels are taken to 10 fb$^{-1}$ with $\sqrt{\mathcal{L}}$ rescaling). Both the Gaussian scaling of the statistical significance and the 2xATLAS approximation are expected to preserve the qualitative features of the expected reach. For illustration, we also show results with 10 and 15 fb$^{-1}$/channel/experiment. The channels all include improvement factors as detailed in the ATLAS note~\cite{atlasnote}.  In our plots we study the expected reach on the $(m_A,\tan\beta)$ plane, fixing the values of the soft parameters. At each point on the plane the Higgs spectrum, decay rates, and production cross sections are calculated with FeynHiggs~\cite{Frank:2006yh,Heinemeyer:1998np,Heinemeyer:1998yj}, and a quantity $R^{95}$ is calculated for each channel as the ratio of the signal (cross section times Higgs decay branching ratio) that can be probed at the 95\% confidence level relative to the signal predicted by the MSSM at that point. We combine the $R^{95}$ values from multiple channels in inverse quadrature, and the LHC is expected to have exclusion power for a point when the combined $R^{95}\leq 1$. More generally, $R^{95}$ is related to the expected statistical significance $\sigma$ of discovery or exclusion via $\sigma \approx 2/R^{95}$. In practice, the inverse quadrature combination results in a reach for the SM-like Higgs boson that is 10-20\% more conservative than the more precise combination performed by ATLAS~\cite{atlasnote}. To compensate, we apply an additional 15\% improvement factor.  When relevant, we also show contours of 3 and 5$\sigma$ reach.  We expect the results presented in this analysis to give a good estimate of the MSSM bounds that would follow for a more precise combination of the ATLAS and CMS results.
%due to the magnitude of the effects discussed in this article, it is 

Similarly to the case of the LHC,
for the analysis involving Tevatron data we estimate the reach by doubling the luminosity at the CDF experiment  with a luminosity of 10~fb$^{-1}$~\cite{cdfnote}  (all channels are taken to 20 fb$^{-1}$ with $\sqrt{\mathcal{L}}$ rescaling.)  We include a 30\% improvement factor to account for ongoing analysis optimizations~\cite{fishertalk} and as performed for the LHC, the $R^{95}$ values from each channel are combined in inverse quadrature. In practice this combination is simple but effective for the Tevatron, differing for the SM Higgs by no more than about 6\% from the results reported in Ref.~\cite{cdfnote}. in the low mass region, with a mean deviation of less than a percent between 110 and 130 GeV. 

\section{The LHC MSSM Higgs Reach}

\begin{figure*}
\begin{minipage}{1.0\linewidth}
\begin{center}
\begin{tabular}{cc}
\includegraphics[width=0.50\textwidth]{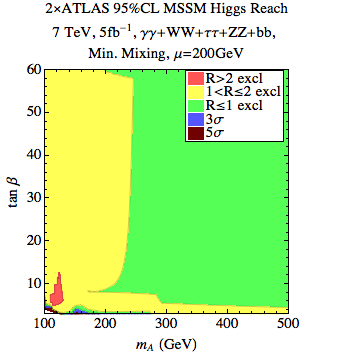} &
\includegraphics[width=0.50\textwidth]{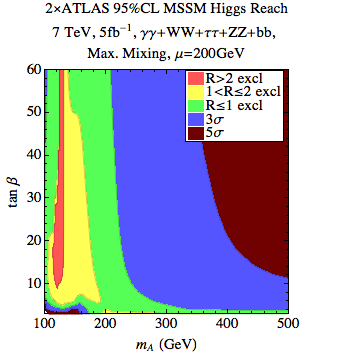} \\
\includegraphics[width=0.50\textwidth]{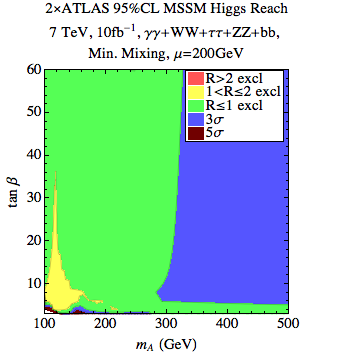} &
\includegraphics[width=0.50\textwidth]{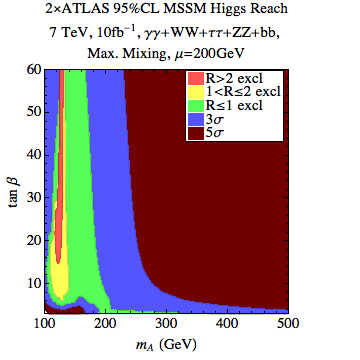}\\
\includegraphics[width=0.50\textwidth]{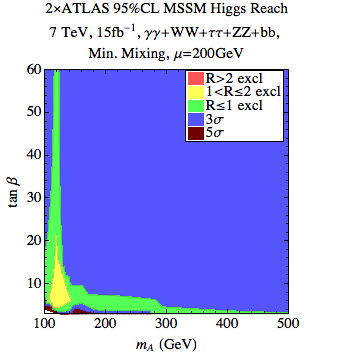} &
\includegraphics[width=0.50\textwidth]{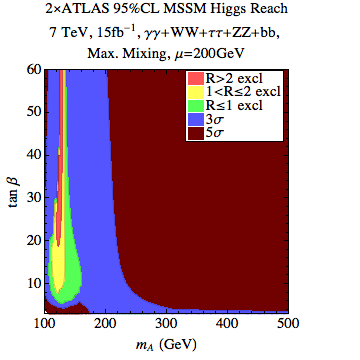} 
\end{tabular}
\caption{Top row: Estimated median LHC reach for the light, SM-like Higgs boson in the minimal mixing (\textit{left}) and maximal mixing (\textit{right}) benchmark scenarios of the MSSM with 5 fb$^{-1}$/experiment. Middle (Bottom) row: same, with 10 (15) fb$^{-1}$/experiment.} 
\label{LHCcombfig}
\end{center}
\end{minipage}
\end{figure*}

We consider first two standard benchmark scenarios, known as the maximal and minimal mixing scenarios~\cite{Carena:2002qg}, for the low-scale soft supersymmetry breaking parameters. The reach for the SM-like Higgs boson is shown in these two models on the $(m_A,\tan\beta)$ plane in Fig.~\ref{LHCcombfig}. For illustration, we give the results for 5, 10, and 15 fb$^{-1}$ of data per experiment.

The projected LHC reach is generally weaker in minimal mixing due to the smaller values of $m_h$ and stronger in maximal mixing where $m_h$ is larger.
For moderate values of $\tan\beta$ and $m_A \gsim 150$~GeV, we obtain $m_h \sim 115-120$~GeV in the minimal mixing scenario and $m_h\sim 120-130$~GeV  in the maximal mixing case. A sizable impact is had by the $h\rightarrow WW$ channel, for which projections were not provided by ATLAS below $m_h=120$ GeV, and is thus absent in the minimal mixing plots.  For low $m_A$, however, the vector boson fusion channel with $h\rightarrow\tau\tau$ and the associated production channel $Vh\rightarrow b\bar{b}$ provide some reach in minimal mixing. Both of these channels grow stronger with smaller $m_h$, so the coverage in this region is stronger than in maximal mixing. In both models, however, it is clear that overall the total reach is suppressed as $m_A$ decreases. As mentioned in the introduction, this is due to tree-level mixing between the $CP$-even Higgs bosons, which can result in an enhanced decay width of the SM-like Higgs into bottom quarks.  Such mixing is stronger for low values of the non-standard Higgs boson masses and tends to suppress the
Higgs decay into photons and $W$ bosons, rendering the searches at the LHC more challenging\footnote{Note that although the $h\rightarrow b\bar{b}$ partial width can easily increase by an order of magnitude, since it is the dominant contributor to the total Higgs width, the $h\rightarrow b\bar{b}$ branching fraction is only increased by a factor $\lesssim 2$. For this reason $Vh\rightarrow b\bar{b}$ does not compensate for the $h\rightarrow\gamma\gamma,WW$ channels, where the branching ratios can experience the full order of magnitude suppression.}.  
%For moderate values of $\tan\beta$ this property is present in both the maximal and minimal mixing scenarios.  
%The enhancement of the SM-like Higgs bottom quark coupling is in general induced by a small mixing between standard and non-standard Higgses, which is dominated by tree-level effects.

\begin{figure*}
\begin{minipage}{1.0\linewidth}
\begin{center}
\begin{tabular}{cc}
\includegraphics[width=0.50\textwidth]{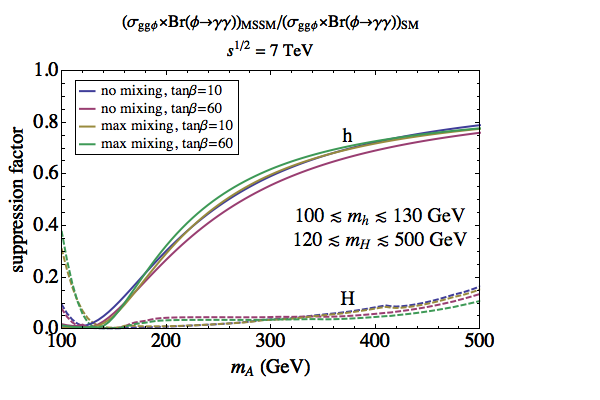} &
\includegraphics[width=0.50\textwidth]{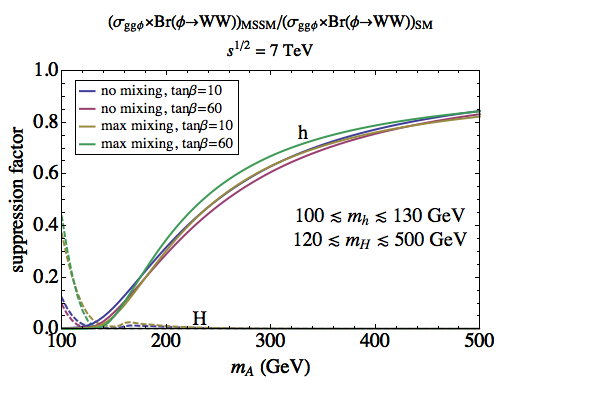} 
\end{tabular}
\caption{Rates for $gg\rightarrow h\rightarrow\gamma\gamma,WW$ (solid) and $gg\rightarrow H\rightarrow\gamma\gamma,WW$ (dashed) in the MSSM, relative to the rates in the SM for a Higgs of mass $m_h$ or $m_H$, respectively. Four different curves are shown for each particle, demonstrating the relatively model-independent nature of the suppression.} 
\label{LHCsuppfig}
\end{center}
\end{minipage}
\end{figure*}

The Higgs doublet mixing decreases as $\cot\beta$ for large values of $\tan\beta$, but since the coupling of the non-standard Higgs
bosons to bottom-quarks is approximately proportional to $\tan\beta$ for large values of $\tan\beta$, the mixing effects on the $BR(h \to b \bar{b})$ remain approximately constant.
%independent of $\tan\beta$. 
This property, as well as the overall magnitude of the suppression effect on the rare decays, is demonstrated in Fig.~\ref{LHCsuppfig} for the $gg\rightarrow h\rightarrow\gamma\gamma,WW$
channels.  We also display the suppression relative to the SM for the $gg\rightarrow H\rightarrow\gamma\gamma,WW$
channels, since below $m_A\sim 130$ GeV the heavy Higgs becomes SM-like in its coupling to gauge bosons, while $h$ becomes nonstandard. However, $H$ still retains an enhanced coupling to $b\bar{b}$ due to a small mixing with $H^0_d$, leading again to a suppression of the $H\rightarrow\gamma\gamma,WW$ rates.

The $b\bar{b}$ enhancement has relevant consequences for searches at the LHC. For 
maximal mixing, in which the SM-Higgs mass is close to 130~GeV, the most important search channel is the decay into a pair
of $W$-gauge bosons. This decay channel is suppressed for small $m_A$. As shown in Fig.~\ref{LHCcombfig},  combining the two LHC
experiments at 5~fb$^{-1}$, for $m_A$ below 200~GeV there are sizable regions where the LHC is not expected to probe the presence of a SM-like Higgs boson
in the standard Higgs search channels.  

In the minimal mixing scenario, the SM-like Higgs boson has a mass close to 115~GeV and the main decay channels are therefore into $\tau$-leptons and $b$-quarks.
The main searches at the LHC are through the Higgs decay into two photons, which as shown in Fig.~\ref{LHCsuppfig}, is strongly suppressed for $CP$-odd Higgs masses
$m_A < 300$~GeV. 

Consequently, in both scenarios, the searches for a MSSM light SM-like Higgs boson at the LHC will depend critically on the performances of the VBF,$h\rightarrow\tau\tau$ and $Vh\rightarrow b\bar{b}$ modes for low values of $m_A$.  The reach for SM-like Higgs bosons in these two channels improves for  smaller values of the Higgs mass, and in combination with the $WW$ and $\gamma\gamma$ channels, we find that the LHC can test the low $m_A$ region in both scenarios at the 2$\sigma$ level with 10~fb$^{-1}$ and find 3$\sigma$ evidence of the SM-like Higgs boson at 15~fb$^{-1}$ in the majority of the low $m_A$ parameter space. 

\begin{figure*}
\begin{minipage}{1.0\linewidth}
\begin{center}
\begin{tabular}{cc}
\includegraphics[width=0.50\textwidth]{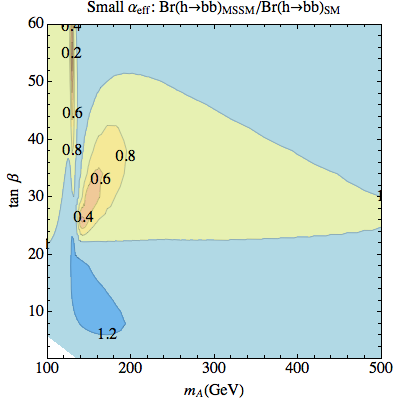} &
\includegraphics[width=0.50\textwidth]{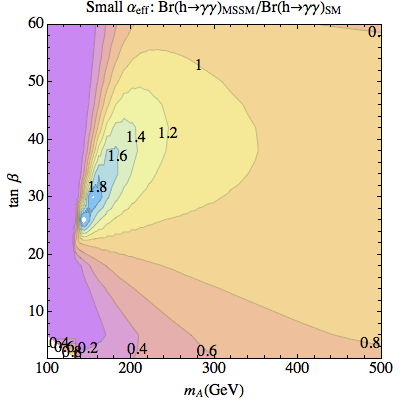} 
\end{tabular}
\caption{Enhancement of the $h \to b b$ decay branching ratio (left panel) and enhancement of the $h \to \gamma\gamma$ decay branching ratio (right panel) in the small
$\alpha$ scenario of the MSSM.} 
\label{hbbgagafig}
\end{center}
\end{minipage}
\end{figure*}

On the other hand, it is also possible to achieve an \textit{increase} in the rates for the $h\rightarrow\gamma\gamma,WW$ decay channels sought at the LHC.  Such effects are also achieved through Higgs mixing: for sufficiently large values of $\tan\beta$, one-loop corrections to the mixing angle may be as important as the tree-level effects. Indeed, the enhancement of the
bottom-quark coupling may be avoided in limited regions of parameter space, in which the stop mixing parameter $A_t$
and the Higgsino mass parameter $\mu$ are larger than the characteristic stop mass scale. For negative values of the product $\mu A_t$, the one-loop corrections
may cancel the tree-level mixing effects, and the SM-like Higgs boson becomes almost purely $H_u^0$. Under these conditions, a large suppression
of the b-quark coupling of the SM-like Higgs may be obtained\footnote{Similar suppression occurs for the $h\tau\tau$ coupling, although for slightly different values of the parameters due to large quantum corrections to the $hb\bar{b}$ coupling that are absent for the $\tau$. This would also suppress the VBF,$h\rightarrow\tau\tau$ channel.}. This possibility has been named small $\alpha_{eff}$ scenario~\cite{Carena:2002qg}, since the fraction of the SM-like Higgs composed of the neutral field coupling to down-quarks and leptons is small.  As shown in Fig.~\ref{hbbgagafig}, the suppression of the Higgs decay width leads to an enhancement of the photon branching ratio.

\begin{figure*}
\begin{minipage}{1.0\linewidth}
\begin{center}
\begin{tabular}{cc}
\includegraphics[width=0.50\textwidth]{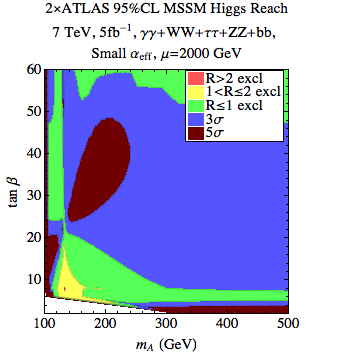}  &
\includegraphics[width=0.50\textwidth]{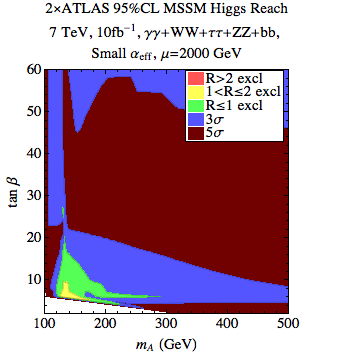}
\end{tabular}
\caption{LHC reach for the light, SM-like Higgs boson in the small $\alpha_{eff}$ benchmark scenario of the MSSM. Left: 5 fb$^{-1}$/experiment; Right: 10 fb$^{-1}$/experiment.}
\label{LHCsmallalphacombfig}
\end{center}
\end{minipage}
\end{figure*}

In the region of parameters where the $h \to \gamma \gamma$ decay rate is enhanced, the LHC has the possibility of a 5$\sigma$ discovery in the near future.
This is shown in the left panel of Fig.~\ref{LHCsmallalphacombfig} for 5 fb$^{-1}$/experiment, and in the right panel for 10 fb$^{-1}$/experiment.

\section{Combination with other Higgs searches at the Tevatron and the LHC}

\begin{figure*}
\begin{minipage}{1.0\linewidth}
\begin{center}
\begin{tabular}{cc}
\includegraphics[width=0.50\textwidth]{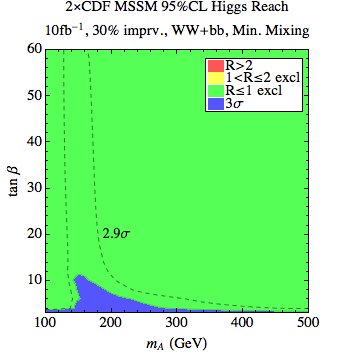} &
\includegraphics[width=0.50\textwidth]{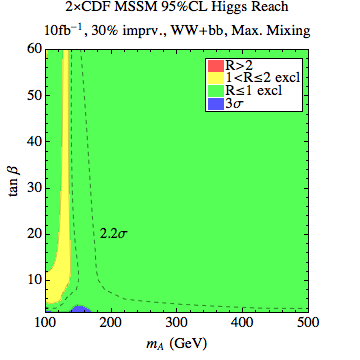} 
\end{tabular}
\caption{Estimated median Tevatron reach for the light, SM-like Higgs boson in the minimal mixing (\textit{left}) and maximal mixing (\textit{right}) benchmark scenarios of the MSSM.} 
\label{Tevcombfig}
\end{center}
\end{minipage}
\end{figure*}

In the last section, we showed that  for low values of $m_A$ searches for the SM-like Higgs boson at the LHC may become challenging.  
We consider two routes to covering the low $m_A$ parameter space. First, because the main Tevatron search mode for light Higgs states is through $h\rightarrow b\bar{b}$ decays, a statistical combination of the datasets may be well-motivated. In Fig.~\ref{Tevcombfig}, we give the estimated Tevatron reach in maximal and minimal mixing. It is clear that the Tevatron should have nearly full exclusion coverage of the MSSM Higgs sector by the time it shuts down. In fact,
%For larger values of $m_A$ the Tevatron is slightly weaker because the Higgs mass increases and the rate into bottom quarks is not enhanced over the SM 
for low values of $m_A$ the Tevatron has of order 20\% greater reach than at large $m_A$, because as $m_A$ decreases the Higgs mass is reduced and the rate into bottom quarks increases. (This feature is not visible in the colors of Fig.~\ref{Tevcombfig} because of the coarse contours in $R^{95}$, which we chose for consistency with the previous LHC figures. To illustrate the behavior we add dashed contours at lower values of $m_A$, inside of which the Tevatron power is higher.)
%However, for low values of $m_A$, the Tevatron is stronger.  
In any case, Fig.~\ref{Tevcombfig} demonstrates that searches for the SM-like Higgs at the Tevatron and the LHC become of similar power and complementary in the early LHC phase.
Therefore, it is worth considering the utility of combining the analyses of the data from both colliders.  
%In order to understand the relevance of this complementary reach, we have performed a combination of the statistical significance of the signals at both colliders.  

\begin{figure*}
\begin{minipage}{1.0\linewidth}
\begin{center}
\begin{tabular}{cc}
\includegraphics[width=0.50\textwidth]{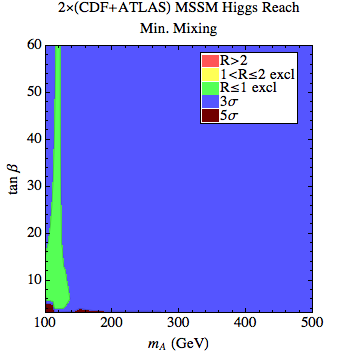} &
\includegraphics[width=0.50\textwidth]{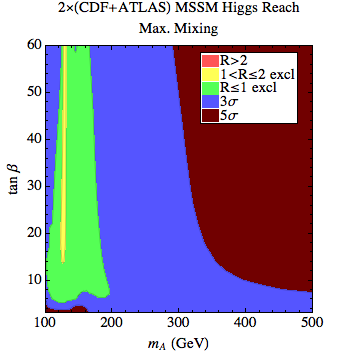} \\
\includegraphics[width=0.50\textwidth]{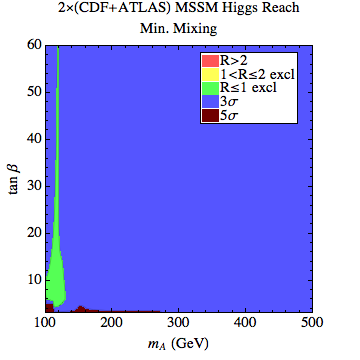} &
\includegraphics[width=0.50\textwidth]{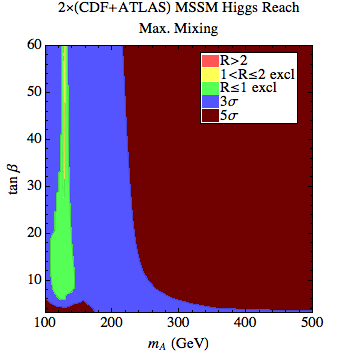}
\end{tabular}
\caption{Estimated median combined Tevatron+LHC reach for the light, SM-like Higgs boson in the minimal mixing (\textit{left}) and maximal mixing (\textit{right}) benchmark scenarios of the MSSM. Top: 5 fb$^{-1}$/experiment for the LHC, 10 fb$^{-1}$/experiment for the Tevatron; Bottom: 10 fb$^{-1}$/experiment for both the Tevatron and LHC.} 
\label{LHCTevfig}
\end{center}
\end{minipage}
\end{figure*}

%In order to quantify the relevance of the complementarity, we show i
In Fig.~\ref{LHCTevfig} we show the combination of
the estimated reaches of the LHC and the Tevatron, using 5 fb$^{-1}$/experiment for the LHC. Most notably the combination leads to evidence for a SM-like Higgs in both the minimal mixing as well as the maximal mixing scenario for most of the parameter space, including the low $m_A$ region.  This stresses the importance
of achieving the efficiency improvements on the search for SM-like Higgs bosons at the Tevatron, and suggests that an effort to perform a combination of the
data from the four experiments after the first year of LHC running is justified.

% This result is shown in Fig.~\ref{LHCcombfig}.  As shown in Fig.~\ref{Tevcombfig}, for such low values of the SM-like Higgs mass, the Tevatron has excellent reach capabilities in most of the parameter space.

%From the above discussion, 

A second approach to studying the low $m_A$ parameter space is given by the LHC searches for the nonstandard Higgs bosons $H$ and $A$ in their decays to $\tau$ leptons. These channels are most effective at low $m_A$, where both $H$ and $A$ are lighter and easier to produce, and at large $\tan\beta$ where the production in association with bottom quarks is proportional to $\tan^2\beta$. 

\begin{figure*}
\begin{minipage}{1.0\linewidth}
\begin{center}
\begin{tabular}{cc}
\includegraphics[width=0.50\textwidth]{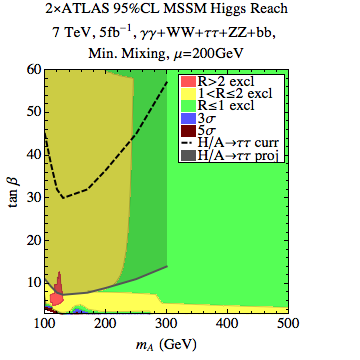} &
\includegraphics[width=0.50\textwidth]{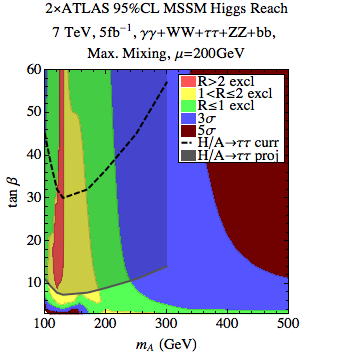} 
\end{tabular}
\caption{LHC reach for the light, SM-like Higgs boson and the nonstandard Higgs states in the minimal mixing (\textit{left}) and maximal mixing (\textit{right}) benchmark scenarios of the MSSM.} 
\label{LHCNSfig}
\end{center}
\end{minipage}
\end{figure*}

In Fig.~\ref{LHCNSfig} we overlay the estimated reach for the neutral MSSM Higgs bosons with nonstandard gauge couplings in the maximal and minimal mixing scenarios. The 95\% CL limit is derived from the expected limits given in the recent ATLAS $H/A\rightarrow \tau\tau$ search with 36 pb$^{-1}$ (which are cut at $m_A=300$ GeV), using the tree-level approximation that the reach in $\tan\beta$ scales like $\mathcal{L}^{1/4}$ and the useful property that the nonstandard Higgs expected reach is robust against changes in the soft parameters~\cite{Carena:2005ek} (although some weak dependence on $\mu$ can appear for large values of $\mu$~\cite{Gennai:2007ys}.) This demonstrates the complementarity of the two types of Higgs searches at the LHC: a statistical combination of the channels should be able to test the parameter space of the model, even though none of the particles $h,H,A$ can necessarily be probed on all of the model space.

\begin{figure*}
\begin{minipage}{1.0\linewidth}
\begin{center}
\includegraphics[width=0.50\textwidth]{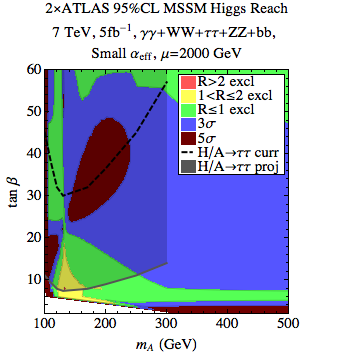} 
\caption{
Same as Fig.~\ref{LHCsmallalphacombfig}, but with nonstandard searches overlaid, showing both the current limits from $H/A\rightarrow \tau\tau$ (dashed curve) and the projected reach with 5 fb$^{-1}$ (shaded region).} 
\label{LHCsmallalphacombNSfig}
\end{center}
\end{minipage}
\end{figure*}

In the regions of parameter space for which the SM-like Higgs bottom and tau couplings are suppressed, analyzed in the small-$\alpha_{eff}$ scenario of Fig.~\ref{LHCsmallalphacombfig}, the LHC will also be able to test the nonstandard Higgs sector. 
%decay production, via the neutral $CP$-even and $CP$-odd Higgs decays into a pair of $\tau$-leptons. 
This is shown in Fig.~\ref{LHCsmallalphacombNSfig}, where the current 95\% CL limit on the $CP$-odd Higgs mass is drawn as a dashed line.  For the specific point we analyzed, the current bounds already heavily
constrain the region of parameters for which the branching ratio $BR(h \to \gamma\gamma)$ may be enhanced. This is a generic feature. In Fig.~\ref{LHCsmallalphacombNSfig} we
also show the projected reach of the $H/A\rightarrow \tau\tau$ channel with 5~fb$^{-1}$ per experiment. Based on these results, we find that with the acquisition of 5~fb$^{-1}$/experiment, either the LHC will find both the SM-like Higgs and evidence of non-standard Higgs bosons, or the region in which the photon pair production is enhanced will be ruled out by both channels.

\section{Conclusions}

In this article we have analyzed the 7 TeV LHC capabilities to exclude, provide evidence for, or discover neutral Higgs bosons in the MSSM. At $m_A\gtrsim 300$ GeV, in the maximal mixing scenario, for which the Higgs mass is about 125-130~GeV, the LHC is expected to discover or find evidence of a SM-like Higgs boson (the state provided by the doublet that is primarily responsible for electroweak symmetry breaking) in a combination of the $WW$ and $\gamma\gamma$ channels with 5~fb$^{-1}$/experiment. In the same region of $m_A$, evidence for $h$ is expected in the diphoton channel with $\gtrsim 10$~fb$^{-1}$/experiment in the minimal mixing scenario, for which the Higgs mass is about 115-120 GeV.  At lower values of $m_A$, we have emphasized that the SM-like Higgs can generically exhibit branching ratios different from those of the SM Higgs in decays relevant for the main LHC search channels. In the most generic models for the soft parameters, the $h\rightarrow\gamma\gamma,WW$ modes are suppressed at low to moderate $m_A$ by a large increase in the $h\rightarrow b\bar{b}$ width, an effect that is due to mixing between the two Higgs doublets. In such cases we have shown that combinations with Tevatron results and with nonstandard Higgs boson searches at the LHC can provide an experimental handle on the parameter space. Furthermore, with other specific choices of the soft parameters, the mixing can be such that the $h\rightarrow b\bar{b}$ width is strongly suppressed, leading to an enhancement in the $h\rightarrow \gamma\gamma,WW$ branching ratios, allowing the discovery of the SM-like Higgs at 5~fb$^{-1}$. Because this feature is present at low $m_A$ and large $\tan\beta$ such models will be probed in the near future by the searches for standard and nonstandard Higgs bosons.

%%%%%%%%%%%%%%%%%%%%%%%%%%%%%%%%%%%%%%%%%%%%%%%%%%%%%%%%%%%%%%%%%%%%%%%%%%%%%%
%%%%%%%%%%%%%%%%%%%%%%%%%%%%%%%%%%%%%%%%%%%%%%%%%%%%%%%%%%%%%%%%%%%%%%%%%%%%%%

%%%%%%%%%%%%%%%%%%%%%%%%%%%%%%%%%%%%%%%%%%%%%%%%%%%%%%%%%%%%%%%%%%%%%%%%%%%%%%
%%%%%%%%%%%%%%%%%%%%%%%%%%%%%%%%%%%%%%%%%%%%%%%%%%%%%%%%%%%%%%%%%%%%%%%%%%%%%%
~\\
{\bf \large Acknowledgements} \\

Fermilab is operated by Fermi Research Alliance, LLC under Contract No. DE-AC02-07CH11359 with the U.S. Department of Energy. Work at ANL is supported in part by the U.S. Department of Energy (DOE), Div.~of HEP, Contract DE-AC02-06CH11357. This work was supported in part by the DOE under Task TeV of contract DE-FGO2-96-ER40956. The work of T.L. was supported by the Department of Energy under
contract DE-FG02-91ER40618.

%\newpage

\FloatBarrier

\end{document}